\documentclass[american,aps,pra,reprint,longbibliography]{revtex4-2}
\usepackage{lmodern}
\usepackage[T1]{fontenc}
\usepackage[latin9]{inputenc}
\usepackage{color}
\usepackage{babel}
\usepackage{float}
\usepackage{amsmath}
\usepackage{amssymb}
\usepackage{mathdots}
\usepackage{graphicx}
\usepackage{esint}
\usepackage[unicode=true,pdfusetitle,
 bookmarks=true,bookmarksnumbered=false,bookmarksopen=false,
 breaklinks=false,pdfborder={0 0 0},pdfborderstyle={},backref=false,colorlinks=true]
 {hyperref}
\hypersetup{
 citecolor=blue,linkcolor=teal,urlcolor=blue}

\makeatletter
\usepackage{xcolor}

\makeatother

\begin{document}
\global\long\def\i{\mathrm{i}}%
\global\long\def\d{\mathrm{d}}%
\global\long\def\e{\mathrm{e}}%
\global\long\def\Tr{\mathrm{Tr}}%
\global\long\def\ad{\text{{\rm ad}}}%
\global\long\def\Ad{\text{{\rm Ad}}}%
\global\long\def\bra#1{\langle#1|}%
\global\long\def\ket#1{|#1\rangle}%
\global\long\def\braket#1#2{\left\langle #1|#2\right\rangle }%
\global\long\def\ketbra#1#2{\left|#1\right\rangle \!\left\langle #2\right|}%
\global\long\def\dbra#1{\langle\!\langle#1|}%
\global\long\def\dket#1{|#1\rangle\!\rangle}%
\global\long\def\dbraket#1#2{\left\langle \!\left\langle #1|#2\right\rangle \!\right\rangle }%
\global\long\def\dketbra#1#2{\left.\!\left|#1\right\rangle \!\right\rangle \!\left\langle \!\left\langle #2\right|\!\right.}%
\global\long\def\ps#1#2#3{\prescript{#1}{#2}{#3}}%

\title{Extended degenerate perturbation theory for the Floquet--Hilbert
space}
\author{Yakov Braver}
\affiliation{Institute of Theoretical Physics and Astronomy, Vilnius University,
Saul\.{e}tekio 3, LT-10257 Vilnius, Lithuania}
\author{Egidijus Anisimovas}
\affiliation{Institute of Theoretical Physics and Astronomy, Vilnius University,
Saul\.{e}tekio 3, LT-10257 Vilnius, Lithuania}
\begin{abstract}
We consider construction of effective Hamiltonians for periodically
driven interacting systems in the case of resonant driving. The standard
high-frequency expansion is not expected to converge due to the resonant
creation of collective excitations, and one option is to resort to
the application of degenerate perturbation theory (DPT) in the Floquet--Hilbert
space. We propose an extension of DPT whereby the degenerate subspace
includes not only the degenerate levels of interest but rather all
levels in a Floquet zone. The resulting approach, which we call extended
DPT (EDPT), is shown to resemble a high-frequency expansion, provided
the quasienergy matrix is constructed such that each $m$th diagonal
block contains energies reduced to the $m$th Floquet zone. The proposed
theory is applied to a driven Bose--Hubbard model and is shown to
yield more accurate quasienergy spectra than the conventional DPT.
The computational complexity of EDPT is intermediate between DPT and
the numerically exact approach, thus providing a practical compromise
between accuracy and efficiency.
\end{abstract}
\maketitle

\section{Introduction}

Intriguing dynamical quantum many-body effects such as prethermalization
\citep{Bukov2015a,Abanin2017,Singh2019,RubioAbadal2020,Fleckenstein2021,Ho2023},
localization \citep{Dunlap1986,Eckardt2009,Bordia2017,Abanin2019},
as well as emergence of topological states \citep{Kitagawa2010,Grushin2014,jotzu14haldane,Mikami2016,Price17PRA}
and discrete time crystals \citep{Sacha2015,Else2016,Yao2017,Zhang2017,Choi2017,Else2020}
have been predicted and realized experimentally in periodically driven
quantum systems. From a theoretical point of view, periodicity of
the drive allows one to employ the Floquet theory \citep{Shirley1965,Sambe1973}
and construct an effective time-independent Hamiltonian $\hat{W}$
that stroboscopically characterizes dynamics of the system \citep{Goldman2014,Bukov2015,Eckardt2017}.
In practice, calculation of effective Hamiltonians has to be carried
out perturbatively. Various high-frequency expansions have been devised
to that end \citep{Rahav2003,Eckardt2015,Bukov2015,Itin2015,Mikami2016,Rodriguez2018},
whereby $\hat{W}$ is constructed as an expansion in powers of $1/\omega$,
with $\omega$ the driving frequency; the low-frequency limit $\omega\to0$
has also been investigated \citep{Rodriguez2018}. The present work
is devoted to the case of resonant driving. Let us introduce the relevant
formalism to set the stage for the presentation of the problem.

Construction of effective Hamiltonians may be conveniently approached
in the extended Floquet--Hilbert space, where time-dependent operators
become infinite matrices that possess a block-banded structure. By
Floquet theorem, the Schr\"{o}dinger equation $\i\partial_{t}\ket{\psi(t)}=\hat{H}(t)\ket{\psi(t)}$
with time-periodic Hamiltonian $\hat{H}(t+T)=\hat{H}(t)$ has solutions
of the form $\ket{\psi_{n}(t)}=\e^{-\i\varepsilon_{n}t}\ket{u_{n}(t)}$,
where $\varepsilon_{n}$ are quasienergies, while $\ket{u_{n}(t+T)}=\ket{u_{n}(t)}$
are periodic functions called Floquet modes (we set $\hbar=1$). The
evolution equation for these functions takes the form $\hat{Q}\ket{u_{n}(t)}=\varepsilon_{n}\ket{u_{n}(t)}$,
where $\hat{Q}(t)=\hat{H}(t)-\i\partial_{t}$ is the quasienergy operator.
The Floquet modes $\ket{u_{n}(t)}$ of the Hilbert space can be regarded
as the elements of the composite Floquet--Hilbert space ${\cal F}$
defined as the direct product of the Hilbert space and the space of
time-periodic functions. We adopt the notation $\dket{u_{n}}$ for
the elements of ${\cal F}$, where the inner product is defined as
$\dbraket{u_{n}}{u_{m}}=\frac{1}{T}\intop_{0}^{T}\braket{u_{n}(t)}{u_{m}(t)}\d t$
\citep{Shirley1965,Sambe1973,Eckardt2015,Rodriguez2018}. The basis
spanning ${\cal F}$ is given by $\dket{\alpha m}\Leftrightarrow\ket{\alpha}\e^{\i m\omega t}$,
with $\ket{\alpha}$'s forming a basis in the Hilbert space. The quasienergy
operator then assumes a block-banded form: 
\begin{equation}
\dbra{\alpha'm'}\bar{Q}\dket{\alpha m}=\bra{\alpha'}\hat{H}_{m'-m}\ket{\alpha}+\delta_{m'm}\delta_{\alpha'\alpha}m\omega\label{eq:usual}
\end{equation}
with the blocks containing the Fourier images $\hat{H}_{m}=\frac{1}{T}\int_{0}^{T}\hat{H}(t)\e^{\i m\omega t}\d t$.
Hereafter we indicate the operators acting in ${\cal F}$ by bars.
The above equation can be illustrated by the following matrix:
\begin{equation}
\bar{Q}=\begin{pmatrix}\ddots & \vdots & \vdots & \vdots & \iddots\\
\cdots & \hat{H}_{0}-\hat{I}\omega & \hat{H}_{-1} & \hat{H}_{-2} & \cdots\\
\cdots & \hat{H}_{1} & \hat{H}_{0} & \hat{H}_{-1} & \cdots\\
\cdots & \hat{H}_{2} & \hat{H}_{1} & \hat{H}_{0}+\hat{I}\omega & \cdots\\
\iddots & \vdots & \vdots & \vdots & \ddots
\end{pmatrix},\label{eq:Q}
\end{equation}
where $\hat{I}$ is the identity operator of the Hilbert space.

Finding a frame where the Hamiltonian is time-independent is equivalent
to block-diagonalizing $\bar{Q}$ since off-diagonal blocks account
for the time dependence. The perturbative expansion of the transformed
quasienergy operator is given by
\begin{equation}
\e^{-\bar{G}}\bar{Q}\e^{\bar{G}}=\bar{Q}^{(0)}+(\bar{W}_{{\rm D}}^{(1)}+\bar{W}_{{\rm X}}^{(1)})+(\bar{W}_{{\rm D}}^{(2)}+\bar{W}_{{\rm X}}^{(2)})+\ldots\label{eq:expan}
\end{equation}
Here, $\bar{Q}^{(0)}$ is the unperturbed part of $\bar{Q}$, while
indices ``D'' and ``X'' refer to the block-diagonal and block-off-diagonal
parts of the given operator, respectively:
\begin{equation}
\begin{split}\dbra{\alpha'm'}\bar{O}_{{\rm D}}\dket{\alpha m} & =\dbra{\alpha'm'}\bar{O}\dket{\alpha m}\delta_{m'm},\\
\dbra{\alpha'm'}\bar{O}_{{\rm X}}\dket{\alpha m} & =\dbra{\alpha'm'}\bar{O}\dket{\alpha m}(1-\delta_{m'm}).
\end{split}
\end{equation}
The goal is to find the unitary operator $\bar{G}=\bar{G}^{(1)}+\bar{G}^{(2)}+\ldots$
that sets the off-diagonal parts $\bar{W}_{{\rm X}}^{(n)}$ to zero,
up to a given order. The remaining block-diagonal operator $\bar{Q}_{0}+\bar{W}_{{\rm D}}^{(1)}+\bar{W}_{{\rm D}}^{(2)}+\ldots$
will have the structure $\delta_{m'm}(\bra{\alpha'}\hat{W}_{{\rm D}}^{[n]}\ket{\alpha}+\delta_{\alpha'\alpha}m\omega)$,
where $\hat{W}_{{\rm D}}^{[n]}\equiv\hat{W}_{{\rm D}}^{(1)}+\hat{W}_{{\rm D}}^{(2)}+\ldots+\hat{W}_{{\rm D}}^{(n)}$
is the effective Hamiltonian.

Calculation of $\hat{W}_{{\rm D}}^{[n]}$ in the high-frequency limit
relies on the assumption that the energy spectrum of the unperturbed
Hamiltonian is bounded and that its width is much less than the driving
frequency $\omega$. In that case, the unperturbed quasienergy operator
can be split as $\bar{Q}=\bar{Q}_{0}+\bar{H}$ with $-\i\partial_{t}\Leftrightarrow\bar{Q}_{0}=\delta_{m'm}\delta_{\alpha'\alpha}m\omega$
and $\hat{H}(t)\Leftrightarrow\bar{H}$. By assumption, the elements
of $\bar{H}$ are small compared to $\omega$, therefore, $\bar{H}$
can be treated perturbatively. The expansion (\ref{eq:expan}) then
becomes an expansion in powers of $1/\omega$. However, if transitions
resonant with $\omega$ are possible, then one is required to include
the diagonal elements of $\bar{H}$ in the unperturbed part of the
problem, but this introduces degeneracies. Specifically, if the difference
$E_{\beta}-E_{\alpha}$ between two diagonal elements of $\hat{H}_{0}$
is equal (or is close to) $n\omega$, where $n$ is integer, then
the degeneracy $\dbra{\beta(m+n)}\bar{Q}\dket{\beta(m+n)}=\dbra{\alpha m}\bar{Q}\dket{\alpha m}$
makes the perturbation theory divergent. In that case, one can apply
the standard degenerate perturbation theory (see e.g. Ref.~\citep{LL3}),
whereby $\hat{W}_{{\rm D}}^{[n]}$ is constructed by including the
couplings between the degenerate states exactly, and taking into account
all the remaining couplings perturbatively \citep{Eckardt2008,Weinberg2015}.

Our present aim is to extend the degenerate perturbation theory in
the Floquet--Hilbert space to obtain expressions for $\hat{W}_{{\rm D}}^{[n]}$
that ensure higher accuracy of the resulting quasienergy spectrum
both exactly on resonance and in its vicinity. This will be achieved
by including in the degenerate subspace not only the degenerate levels
of interest but rather all the states of the system. The resulting
approach, called EDPT, parallels the van Vleck high-frequency expansion
\citep{Eckardt2015} provided the elements of $\bar{Q}$ are reordered
so that each $m$th diagonal block corresponds to the $m$th Floquet
zone. To demonstrate the validity of the obtained expressions, we
apply them to the calculation of quasienergy spectra of the driven
Bose--Hubbard model for a number of parameter sets. Comparison with
numerically exact results shows that EDPT surpasses the conventional
degenerate perturbation theory in terms of accuracy while requiring
less computational effort than the exact approach.

\section{Degenerate perturbation theories in the extended space}

\begin{figure*}
\begin{centering}
\includegraphics{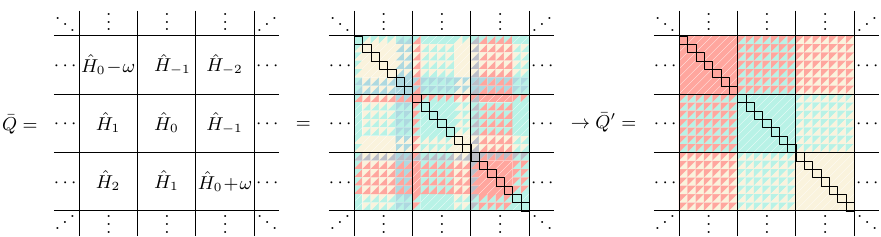}
\par\end{centering}
\caption{\label{fig:scheme}Schematic representation of $\bar{Q}$ and $\bar{Q}'$.
In the first step, the diagonal elements of $\bar{Q}$ are colored
such that those lying in the same Floquet zone share the same color.
For example, the diagonal elements falling in the range $[-\frac{\omega}{2},\frac{\omega}{2})$
are colored red, those falling in the range $[\frac{\omega}{2},\frac{3\omega}{2})$
are colored blue, and so on. Additionally, each off-diagonal element
(in both diagonal and off-diagonal blocks) is colored in two tones
corresponding to the colors of diagonal elements that are being coupled.
In the second step, all elements are permuted so that like-colored
diagonal elements are gathered in the same blocks (while the permutation
of the off-diagonal elements follows unambiguously).}
\end{figure*}
The starting point of the proposed theory is the natural concept of
reduced energies
\begin{equation}
\varepsilon_{\alpha}^{(0)}=E_{\alpha}-a\omega\in\text{FZ},\label{eq:reduced}
\end{equation}
which are the diagonal elements $E_{\alpha}$ of $\hat{H}_{0}$ reduced
to the chosen Floquet zone (FZ), whose width necessarily equals $\omega$.
Note that $\hat{H}_{0}$ is just the unperturbed Hamiltonian with,
possibly, the secular contribution of the driving included. This way,
an integer $a$ is uniquely assigned to each state $\ket{\alpha}$;
generally, multiple states will share the same value of $a$. We reserve
the symbols $a$, $a'$, $b$, and $c$ to indicate the ``reduction
numbers'' of states $\ket{\alpha}$, $\ket{\alpha'}$, $\ket{\beta}$,
and $\ket{\gamma}$, respectively. Next, we reorder the elements of
the quasienergy operator so that its $m$th diagonal blocks contains
energies reduced to the $m$th FZ. The diagonal elements of the resulting
quasienergy operator $\bar{Q}'$ then read
\begin{equation}
\varepsilon_{\alpha m}^{(0)}\equiv\dbra{\alpha m}\bar{Q}'\dket{\alpha m}=\varepsilon_{\alpha}^{(0)}+m\omega,\label{eq:e_am}
\end{equation}
and an arbitrary element of $\bar{Q}'$ is expressed as
\begin{equation}
\dbra{\alpha'm'}\bar{Q}'\dket{\alpha m}=\bra{\alpha'}\hat{H}_{a-a'+m'-m}\ket{\alpha}+\delta_{m'm}\delta_{\alpha'\alpha}m\omega.\label{eq:main}
\end{equation}
The quasienergy matrix retains its block structure, which can be visualized
as follows:
\begin{align}
 & \bar{Q}'=\begin{pmatrix}\ddots & \vdots & \vdots & \vdots & \iddots\\
\cdots & \hat{D}-\hat{I}\omega & \hat{X}_{-1} & \hat{X}_{-2} & \cdots\\
\cdots & \hat{X}_{1} & \hat{D} & \hat{X}_{-1} & \cdots\\
\cdots & \hat{X}_{2} & \hat{X}_{1} & \hat{D}+\hat{I}\omega & \cdots\\
\iddots & \vdots & \vdots & \vdots & \ddots
\end{pmatrix},\nonumber \\
 & \hat{D}=\begin{pmatrix}\varepsilon_{\alpha}^{(0)} & H_{b-a}^{\alpha\beta} & H_{c-a}^{\alpha\gamma}\\
H_{a-b}^{\beta\alpha} & \varepsilon_{\beta}^{(0)} & H_{c-b}^{\beta\gamma}\\
H_{a-c}^{\gamma\alpha} & H_{b-c}^{\gamma\beta} & \varepsilon_{\gamma}^{(0)}
\end{pmatrix},\label{eq:Q'}\\
 & \hat{X}_{m}=\begin{pmatrix}H_{m}^{\alpha\alpha} & H_{b-a+m}^{\alpha\beta} & H_{c-a+m}^{\alpha\gamma}\\
H_{a-b+m}^{\beta\alpha} & H_{m}^{\beta\beta} & H_{c-b+m}^{\beta\gamma}\\
H_{a-c+m}^{\gamma\alpha} & H_{b-c+m}^{\gamma\beta} & H_{m}^{\gamma\gamma}
\end{pmatrix}.\nonumber 
\end{align}
Here, the matrices $\hat{D}$ and $\hat{X}_{m}$ are shown for the
case of a three-level system for brevity, and $H_{m}^{\alpha'\alpha}\equiv\bra{\alpha'}\hat{H}_{m}\ket{\alpha}$.
The central block $\hat{D}$ describes the states of the central ($m=0$)
FZ and their mutual couplings, which will be accounted for exactly.
The off-diagonal blocks $\hat{X}_{m}$ describe couplings between
different Floquet zones; these couplings will be taken into account
perturbatively. The connection between $\bar{Q}$ in Eq.~(\ref{eq:Q})
and $\bar{Q}'$ in Eq.~(\ref{eq:Q'}) is shown schematically in Fig.~(\ref{fig:scheme}).
The diagonal blocks no longer share common elements, with one possible
exception in case there are states with reduced energies near both
boundaries of the Floquet zones. For example, for the FZ choice $[-\frac{\omega}{2},\frac{\omega}{2})$,
this will be the case if $\varepsilon_{\pm}^{(0)}\approx\pm\frac{\omega}{2}$.
The element $\varepsilon_{+}^{(0)}+m\omega$ of the $m$th diagonal
block will then coincide with the element $\varepsilon_{-}^{(0)}+(m+1)\omega$
of the $(m+1)$st diagonal block. This issue will be discussed further
in Section \ref{subsec:Convergence-condition}. Abstracting from it,
we proceed to derive the expressions for the effective Hamiltonian.

\subsection{Extended degenerate perturbation theory}

In the first step, we separate $\bar{Q}'$ into the unperturbed part
and the perturbation, the latter having a block-diagonal part and
a block-off-diagonal one:
\begin{equation}
\bar{Q}'=\bar{Q}'^{(0)}+\lambda\bar{V}_{{\rm D}}+\lambda\bar{V}_{{\rm X}},
\end{equation}
where
\begin{equation}
\begin{split}\dbra{\alpha'm'}\bar{Q}^{\prime(0)}\dket{\alpha m} & =\varepsilon_{\alpha m}^{(0)}\delta_{\alpha'\alpha}\delta_{m'm},\\
\dbra{\alpha'm'}\bar{V}_{{\rm D}}\dket{\alpha m} & =\bra{\alpha'}\hat{H}_{a-a'}\ket{\alpha}(1-\delta_{\alpha'\alpha})\delta_{m'm},\\
\dbra{\alpha'm'}\bar{V}_{{\rm X}}\dket{\alpha m} & =\bra{\alpha'}\hat{H}_{a-a'+m'-m}\ket{\alpha}(1-\delta_{m'm}).
\end{split}
\label{eq:blocks}
\end{equation}
The dimensionless parameter $\lambda$ has been introduced to track
the order of the expansion. Inserting $\bar{Q}'$ into Eq.~(\ref{eq:expan})
and assuming $\bar{G}=\lambda\bar{G}^{(1)}+\lambda^{2}\bar{G}^{(2)}+\ldots$,
one can collect the terms of the same order. In the first order, this
yields
\begin{equation}
-[\bar{G}^{(1)},\bar{Q}'^{(0)}]+\bar{V}_{{\rm D}}+\bar{V}_{{\rm X}}=\bar{W}_{{\rm D}}^{(1)}+\bar{W}_{{\rm X}}^{(1)}.
\end{equation}
We require $\bar{G}^{(1)}$ be block-off-diagonal, so that $[\bar{G}^{(1)},\bar{Q}'^{(0)}]$
is block-off-diagonal as well. The remaining block-diagonal terms
immediately yield the first-order term of the effective Hamiltonian
according to
\begin{equation}
\dbra{\alpha'm}\bar{W}_{{\rm D}}^{(1)}\dket{\alpha m}=\dbra{\alpha'm}\bar{V}_{{\rm D}}\dket{\alpha m},\label{eq:WD1-F}
\end{equation}
or, in the Hilbert space,
\begin{equation}
\bra{\alpha'}\hat{W}_{{\rm D}}^{(1)}\ket{\alpha}=\bra{\alpha'}\hat{H}_{a-a'}\ket{\alpha}.\label{eq:WD1-H}
\end{equation}

Next, we require $\bar{W}_{{\rm X}}^{(1)}=0$, obtaining the equation
for the block-off-diagonal terms: $[\bar{G}^{(1)},\bar{Q}'^{(0)}]=\bar{V}_{{\rm X}}$.
This allows us to find $\bar{G}^{(1)}$ and proceed to the equation
for the second-order terms. Continuing in the similar fashion, one
obtains
\begin{equation}
\begin{split}\dbra{\alpha'm}\bar{W}_{{\rm D}}^{(2)}\dket{\alpha m} & =\frac{1}{2}\sum_{\beta}\sum_{n\neq m}\dbra{\alpha'm}\bar{V}_{{\rm X}}\dket{\beta n}\dbra{\beta n}\bar{V}_{{\rm X}}\dket{\alpha m}\\
 & \times\left(\frac{1}{\varepsilon_{\alpha'm}^{(0)}-\varepsilon_{\beta n}^{(0)}}+\frac{1}{\varepsilon_{\alpha m}^{(0)}-\varepsilon_{\beta n}^{(0)}}\right).
\end{split}
\label{eq:WD2-F}
\end{equation}
One recognizes that the resulting expressions are identical to the
van Vleck high-frequency expansion \citep{Eckardt2015}, which is
expected since $\bar{Q}'$ possesses exactly the same structure as
in the usual applications of this expansion. The Hilbert-space expressions,
however, do different because the matrix elements of $\bar{V}_{{\rm X}}$
and the quantities $\varepsilon_{\alpha m}^{(0)}$ have a different
meaning in our case. The second-order term of the effective Hamiltonian
results as

\begin{equation}
\begin{split}\bra{\alpha'}\hat{W}_{{\rm D}}^{(2)}\ket{\alpha} & =\frac{1}{2}\sum_{\beta}\sum_{n\neq0}\bra{\alpha'}\hat{H}_{-(a'-b+n)}\ketbra{\beta}{\beta}\hat{H}_{a-b+n}\ket{\alpha}\\
 & \times\left(\frac{1}{\varepsilon_{\alpha'}^{(0)}-\varepsilon_{\beta}^{(0)}-n\omega}+\frac{1}{\varepsilon_{\alpha}^{(0)}-\varepsilon_{\beta}^{(0)}-n\omega}\right).
\end{split}
\label{eq:WD2-H}
\end{equation}
Expressions for the third-order terms of the effective Hamiltonian
are provided in Appendix A.

The formula (\ref{eq:WD2-F}) could alternatively be obtained by a
direct application of the conventional degenerate perturbation theory
(DPT) widely used in the context of ordinary Hilbert space \citep{LL3},
provided one assumes that all of the states $\dket{\alpha m}$ of
a given block ($m=0$, say) constitute the degenerate subspace. The
condition $n\neq m$ in the summation over the intermediate states
$\dket{\beta n}$ in Eq.~(\ref{eq:WD2-F}) corresponds precisely
to the skipping of states belonging to the degenerate subspace. We
will refer to the Eqs.~(\ref{eq:WD1-H}) and (\ref{eq:WD2-H}) as
the results of the \emph{extended degenerate perturbation theory}
(EDPT) since the degenerate subspace is extended to include all levels
in a Floquet zone.

\subsection{Conventional degenerate perturbation theory\label{subsec:DPT}}

For comparison, let us consider the conventional DPT, whereby the
degenerate subspace of ${\cal F}$ consists only of the states that
are exactly (or nearly) degenerate. We start by diving the Hilbert
space ${\cal H}$ into two subspaces, $\mathbb{D}^{0}$ and $\mathbb{D}^{1}$,
so that $\mathbb{D}^{1}$ contains the states sharing the same value
of reduced energy of interest, $\varepsilon_{*}^{(0)}$, while $\mathbb{D}^{0}$
contains the remaining states. That is, for each state $\ket{\alpha}$,
\begin{equation}
\ket{\alpha}\in\begin{cases}
\mathbb{D}^{1}, & \varepsilon_{\alpha}^{(0)}=\varepsilon_{*}^{(0)},\\
\mathbb{D}^{0}, & \varepsilon_{\alpha}^{(0)}\neq\varepsilon_{*}^{(0)}.
\end{cases}
\end{equation}
Consequently, we partition each diagonal block of $\bar{Q}'$ into
two subblocks --- one containing the states belonging to $\mathbb{D}^{0}$,
and one containing the states of $\mathbb{D}^{1}$. The couplings
between these subblocks are then considered to constitute the off-diagonal
blocks of $\bar{Q}'$, and are therefore treated as a perturbation.
This means that the definition of diagonal and off-diagonal blocks
is changed from the one given in Eq.~(\ref{eq:blocks}) to
\begin{equation}
\begin{split}\dbra{\alpha'm'}\bar{O}_{{\rm D}}\dket{\alpha m} & =\dbra{\alpha'm}\bar{O}\dket{\alpha m}\delta_{m'm}\delta_{A'A},\\
\dbra{\alpha'm'}\bar{O}_{{\rm X}}\dket{\alpha m} & =\dbra{\alpha'm'}\bar{O}\dket{\alpha m}(1-\delta_{m'm}\delta_{A'A}).
\end{split}
\end{equation}
Here, $\ket{\alpha}\in\mathbb{D}^{A}$ and $\ket{\alpha'}\in\mathbb{D}^{A'}$
so that the Kronecker delta $\delta_{A'A}$ is unity if both states
$\ket{\alpha'}$ and $\ket{\alpha}$ belong to the same subspace ---
either degenerate or not --- and zero if they belong to different
subspaces. With these definitions, and $\dbra{\alpha'm'}\bar{V}\dket{\alpha m}=\bra{\alpha'}\hat{V}_{a-a'+m'-m}\ket{\alpha}$
as before, one finds
\begin{equation}
\bra{\alpha'}\hat{w}_{{\rm D}}^{(1)}\ket{\alpha}=\delta_{A'A}\bra{\alpha'}\hat{H}_{a-a'}\ket{\alpha}\label{eq:WD1-H-1}
\end{equation}
in the first order and 
\begin{equation}
\begin{split}\bra{\alpha'}\hat{w}_{{\rm D}}^{(2)} & \ket{\alpha}=\frac{1}{2}\delta_{A'A}\\
 & \times\sum_{\beta}\sum_{n\neq0\text{ if }B=A}\bra{\alpha'}\hat{H}_{-(a'-b+n)}\ketbra{\beta}{\beta}\hat{H}_{a-b+n}\ket{\alpha}\\
 & \times\left(\frac{1}{\varepsilon_{\alpha'}^{(0)}-\varepsilon_{\beta}^{(0)}-n\omega}+\frac{1}{\varepsilon_{\alpha}^{(0)}-\varepsilon_{\beta}^{(0)}-n\omega}\right)
\end{split}
\label{eq:WD2-H-DPT}
\end{equation}
in the second. In the last expression, $B$ is the subspace number
which $\ket{\beta}$ belongs to, i.e., $\ket{\beta}\in\mathbb{D}^{B}$.
The terms of the effective Hamiltonian constructed using DPT are denoted
here by $\hat{w}^{(n)}$, and they are composed of two uncoupled blocks.
In fact, when using DPT, one is only interested in the block describing
the degenerate subspace, and it can be diagonalized separately from
the other block.

Notably, construction of $\hat{w}_{{\rm D}}^{[2]}$ is not equivalent
to simply neglecting in $\hat{W}_{{\rm D}}^{(2)}$ the couplings between
the blocks $\mathbb{D}^{0}$ and $\mathbb{D}^{1}$. Considering two
states $\ket{\alpha'}$ and $\ket{\alpha}$ of the degenerate subspace,
one has $\bra{\alpha'}\hat{w}_{{\rm D}}^{(2)}\ket{\alpha}\neq\bra{\alpha'}\hat{W}_{{\rm D}}^{(2)}\ket{\alpha}$.
This point will be discussed further in Section \ref{subsec:case-1}.

The above formalism also allows one to construct schemes that are
intermediate between DPT and EDPT. Instead of including in $\mathbb{D}^{1}$
only the states whose reduced energies are equal to the reduced energy
of interest $\varepsilon_{*}^{(0)}$, one can include all states in
a certain interval $[\varepsilon_{*}^{(0)}-\Delta\varepsilon/2,\varepsilon_{*}^{(0)}+\Delta\varepsilon/2]$
(with $\Delta\varepsilon<\omega$). The size of $\mathbb{D}^{1}$
will then increase, resulting in larger numerical effort required
to calculate the eigenvalues, but these should approximate the exact
values more accurately. The choice $\Delta\varepsilon=\omega$ corresponds
to EDPT, whereby all states are assigned to $\mathbb{D}^{1}$. We
will not, however, consider the accuracy of these intermediate schemes
further in this work, instead focusing on the two limiting cases,
DPT and EDPT.

\subsection{Convergence condition\label{subsec:Convergence-condition}}

It follows that the necessary condition for the convergence of the
expansion (\ref{eq:expan}) is 
\begin{equation}
r\equiv\frac{|\bra{\alpha'}\hat{H}_{a-a'-n}\ket{\alpha}|}{|\varepsilon_{\alpha'}^{(0)}-\varepsilon_{\alpha}^{(0)}-n\omega|}\ll1,\label{eq:r}
\end{equation}
which has to be satisfied for all states $\ket{\alpha}$, $\ket{\alpha'}$
and all integers $n$, except those excluded in the summation in Eq.~(\ref{eq:WD2-H})
(EDPT case) or Eq.~(\ref{eq:WD2-H-DPT}) (DPT case). As long as the
numerator does not vanish, this condition might be violated for $|\varepsilon_{\alpha'}^{(0)}-\varepsilon_{\alpha}^{(0)}|\approx\omega$,
which is possible if the reduced energies of the two states $\ket{\alpha'}$
and $\ket{\alpha}$ are on the opposite boundaries of the FZ. A simple
resolution is to shift the FZ in a such way so that there are no levels
near one or both boundaries. However, if the reduced energies $\varepsilon_{\alpha}^{(0)}$
fill the FZ densely, then the problem cannot be circumvented. The
EDPT will be applicable in those cases only if the couplings between
states on the boundary of the FZ can be disregarded. On the other
hand, in case of DPT this issue does not arise for the states of the
degenerate subspace if one centers the FZ around $\varepsilon_{*}^{(0)}$.
Since $\ket{\alpha}$ and $\ket{\alpha'}$ enumerate only the states
of the degenerate subspace, the absolute values of denominators in
Eq.~(\ref{eq:WD2-H-DPT}) are no smaller than $\omega/2$. However,
the problem appears in the DPT case as well once the third-order term,
provided in Appendix A, is included since it contains differences
between the reduced energies of the nondegenerate subspace.

\section{Applications}

To check the validity of the presented theories, we will compare the
quasienergy spectra obtained by diagonalizing $\hat{W}_{{\rm D}}^{[3]}$
and $\hat{w}_{{\rm D}}^{[3]}$ with the numerically exact ones. The
latter were obtained by diagonalizing the single-period evolution
operator, calculated by propagating the Schr\"{o}dinger equation
\citep{Holthaus2015}. The theory will be applied to a driven Bose--Hubbard
system, defined in Section \ref{subsec:Driven-Bose=002013Hubbard-model}.

\begin{figure*}
\begin{centering}
\includegraphics{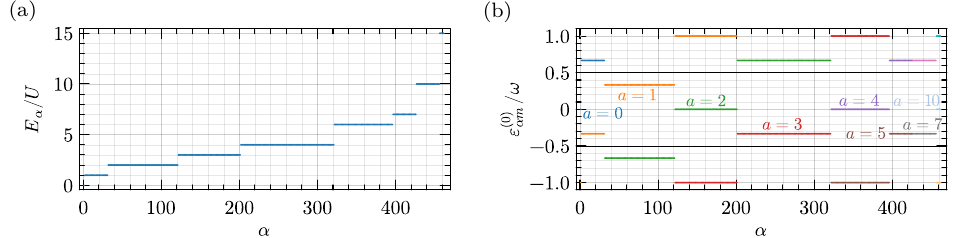}
\par\end{centering}
\caption{\label{fig:1}Diagonal elements of $\hat{H}_{0}$ and $\bar{Q}^{\prime}$
for a BH system on a $1\times6$ lattice assuming $\omega/J=20$,
$U=\frac{2}{3}\omega$. (a) Diagonal elements $E_{\alpha}=\protect\bra{\alpha}\hat{H}_{0}\protect\ket{\alpha}$.
(b) Diagonal elements $\varepsilon_{\alpha m}^{(0)}=\protect\dbra{\alpha m}\bar{Q}^{\prime}\protect\dket{\alpha m}$
in the vicinity of zero. The values of $m$ are: $m=0$ for the levels
in the central FZ $[-\frac{\omega}{2},\frac{\omega}{2})$, $m=-1$
for the levels in the zone below, and $m=1$ for the levels in the
zone above.}
\end{figure*}

\subsection{Driven Bose--Hubbard model\label{subsec:Driven-Bose=002013Hubbard-model}}

The driven Bose--Hubbard model is defined by the Hamiltonian \citep{Eckardt2005}
\begin{equation}
\hat{H}'(t)=-J\sum_{\left\langle ij\right\rangle }\hat{a}_{i}^{\dagger}\hat{a}_{j}+\frac{U}{2}\sum_{j}\hat{n}_{j}(\hat{n}_{j}-1)+\sum_{j}\hat{n}_{j}x_{j}F\cos\omega t
\end{equation}
where $J$, $U$, and $F$ control the strengths of, respectively,
the nearest-neighbor hopping, the on-site interaction, and the external
driving (we study monochromatic driving of frequency $\omega$). The
first sum runs over nearest-neighbor pairs, while the remaining ones
run over all lattice sites. In the last sum, $x_{j}$ is the $x$-coordinate
of site $j$, in units of the lattice constant. A gauge transformation
$\hat{H}(t)=\hat{U}^{\dagger}(t)\hat{H}'(t)\hat{U}(t)-\i\hat{U}^{\dagger}(t)\d_{t}\hat{U}(t)$
with $\hat{U}(t)=\exp\!\left(-\i\frac{F}{\omega}\sin\omega t\sum_{j}x_{j}\hat{n}_{j}\right)$
shows that the effect of the driving amounts to a renormalization
of the hopping strength \citep{Eckardt2017}: 
\begin{equation}
\hat{H}(t)=-J\sum_{\left\langle ij\right\rangle }\e^{\i\frac{F}{\omega}(x_{i}-x_{j})\sin\omega t}\hat{a}_{i}^{\dagger}\hat{a}_{j}+\frac{U}{2}\sum_{j}\hat{n}_{j}(\hat{n}_{j}-1).\label{eq:Ht}
\end{equation}
The Fourier image of the resulting Hamiltonian is given by
\begin{equation}
\hat{H}_{m}=-\sum_{\left\langle ij\right\rangle }J{\cal J}_{m}\!\left(\frac{F(x_{i}-x_{j})}{\omega}\right)\hat{a}_{i}^{\dagger}\hat{a}_{j},\label{eq:Hm}
\end{equation}
where ${\cal J}_{m}(x)$ denotes the Bessel function of the first
kind of order $m$.

We will use the Fock basis $\ket{\alpha}$ to refer to the elements
of $\hat{H}(t)$, denoting the diagonal ones by $E_{\alpha}=\bra{\alpha}\hat{H}_{0}\ket{\alpha}$.
An example of the distribution of diagonal elements $E_{\alpha}$
is displayed in Fig.~\ref{fig:1}(a). The presence of degenerate
elements does not require extra care --- these degeneracies will
directly translate into degeneracies in the first-order term (\ref{eq:WD2-F})
of the effective Hamiltonian. The effective Hamiltonian, once obtained
perturbatively, will be diagonalized exactly (numerically). On the
other hand, the possibility of resonant transitions requires the application
of degenerate perturbation theory in ${\cal F}$. Since the diagonal
elements $E_{\alpha}$ are given by integer multiples of $U$, resonant
transitions become possible when the condition $pU=q\omega$, where
$\{p,q\}\in\mathbb{Z}$, is satisfied. This translates into degeneracies
in ${\cal F}$, as depicted in Fig.~\ref{fig:1}(b) showing the diagonal
elements $\varepsilon_{\alpha m}^{(0)}$ {[}see Eq.~(\ref{eq:e_am}){]}
for $3U=2\omega$. In the figure, black horizontal lines delimit the
FZ chosen as $[-\frac{\omega}{2},\frac{\omega}{2})$, and the values
of $a$ are displayed.

\begin{figure*}
\begin{centering}
\includegraphics{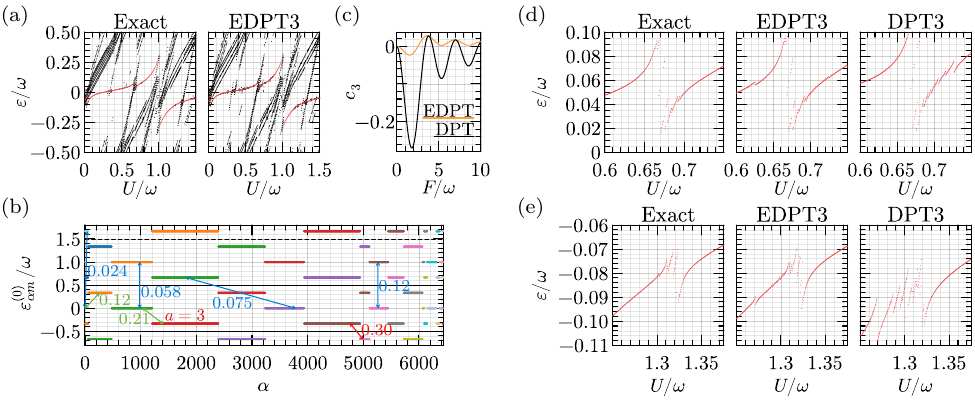}
\par\end{centering}
\caption{\label{fig:2}Assessment of accuracy of DPT and EDPT for an eight-particle
BH system on a $1\times8$ lattice with $F/\omega=2$, $\omega/J=20$.
(a) Quasienergy spectrum for $U\in[0,1.5\omega]$. Quasienergies of
ten Floquet modes having the largest overlap with the MI state are
displayed; quasienergies of the mode with the maximum overlap are
highlighted in red. (b) Diagonal elements of $\bar{Q}'$ at $U=\frac{2}{3}\omega$.
Black horizontal lines indicate the boundaries of the FZs. Red and
blue arrows with numbers show the largest relative coupling strengths
$r$ (\ref{eq:r}) corresponding to coupling with states outside the
central FZ. For example, for the states $\protect\dket{\alpha0}$
of the central FZ with $\alpha\in[478,1205]$, the largest relative
coupling strength of $r=0.058$ is found as a result of the coupling
with one of the states $\protect\dket{\alpha1}$. Green arrows with
numbers show the largest coupling strengths $r$ corresponding to
coupling with states inside the central FZ. Only some of the couplings
are indicated. (c) Coupling element $c_{3}\equiv\protect\bra 3\hat{W}_{{\rm D}}^{(2)}\protect\ket{{\rm MI}}$
calculated at $U=\frac{2}{3}\omega$ using DPT and EDPT versus $F/\omega$.
(d) Quasienergies of the driven MI state in the vicinity of $U=\frac{2}{3}\omega$.
(e) Quasienergies of the driven MI state in the vicinity of $U=\frac{4}{3}\omega$.}
\end{figure*}
The analysis of accuracy of DPT and EDPT will be performed by choosing
the driving strength $F$ and the frequency $\omega$, and calculating
the quasienergies as the parameter $U$ is varied. We will focus our
attention on the quasienergy of the ``driven Mott-insulator (MI)
state'', a name we will use for the Floquet mode having the largest
overlap with MI state (denoted as $\ket{\text{MI}}$). For $U\gg J$,
$\ket{\text{MI}}$ is the ground state of the undriven system, corresponding
to $E=0$. Therefore, we center the FZ around $\varepsilon_{*}^{(0)}=0$
by choosing the interval $[-\frac{\omega}{2},\frac{\omega}{2})$.

We note that the interesting features of the quasienergy spectra ---
the anticrossings indicatory of resonant processes --- will appear
at certain values of the ratio $U/\omega$. From the definition (\ref{eq:reduced})
with $E_{\alpha}=k_{\alpha}U$ where $k_{\alpha}$ is integer, it
follows that the denominator in Eq.~(\ref{eq:r}), $\omega|(\varepsilon_{\alpha'}^{(0)}-\varepsilon_{\alpha}^{(0)})/\omega-n|$,
grows linearly in $\omega$ for fixed $U/\omega$. Therefore, the
accuracy of the perturbation theory in the vicinity of resonances
is expected to increase with increasing $\omega$, although this reasoning
does not take into account the dependence on $\omega$ of the coupling
strength {[}i.e. the numerator in Eq.~(\ref{eq:r}){]}.

Finally, let us clarify that DPT can be applied not only exactly on
resonance (when the degenerate energies of interest exactly coincide),
but also in its vicinity. For example, we can calculate the quasienergies
for $U$ in the vicinity of $\frac{2}{3}\omega$ by including in $\mathbb{D}^{1}$
the same states as those constituting $\mathbb{D}^{1}$ when $U$
exactly equals $\frac{2}{3}\omega$. Similarly, the same reduction
numbers {[}see Eq.~(\ref{eq:reduced}){]} as those obtained exactly
on resonance are used even when $U$ does not exactly equal $\frac{2}{3}\omega$.

\subsection{Study case 1: One-dimensional lattice\label{subsec:case-1}}

We begin the assessment of accuracy of the perturbation theories with
the study of a BH system defined on a periodic $1\times8$ lattice
containing 8 bosons; we set $F/\omega=2$ and $\omega/J=20$. The
plots in Fig.~\ref{fig:2}(a) display the quasienergies of ten Floquet
modes having the largest overlap with the MI state for $U\in[0,1.5\omega]$.
The quasienergies of the driven MI state are highlighted in red. The
left plot displays the exact result, while the right one shows the
results obtained using third-order EDPT. As explained above, the DPT
approach is only applicable in the vicinity of resonances, and cannot
be directly used to calculate the quasienergies for such a wide range
of $U$. The most pronounced anticrossing seen at $U\approx\omega$
corresponds to first-order creation of particle--hole excitations
in the MI state, whereby a particle is annihilated at a certain site
and created at its neighboring site \citep{Eckardt2008}. At $U=\omega$,
transitions between all levels of the system become resonant, therefore,
all diagonal elements of $\bar{Q}'$ in the given diagonal block share
the same value. Consequently, the actual distribution of quasienergies
is almost entirely captured by the first-order effective Hamiltonian
(\ref{eq:WD1-H}). Additionally, DPT becomes equivalent to EDPT since
all levels of the system are included in the degenerate subspace.
The second- and third-order terms of $\hat{W}_{{\rm D}}$ provide
a slight improvement and yield results in agreement with the exact
ones, as shown in Fig.~\ref{fig:2}(a). Notably, the EDPT results
remain sufficiently accurate away from resonances as well. Exactly
on the resonance $U=\omega$, the largest value of the coupling ratio
(\ref{eq:r}) is $r_{\max}=0.12$, which is one of the largest values
that can be considered to satisfy the condition (\ref{eq:r}). Therefore,
for smaller values of $\omega$ (and the same value of $F/\omega$),
the EDPT is not expected to yield reliable results near the resonance
$U=\omega$.

To assess the accuracy of the methods on a finer scale, we inspect
the quasienergy of the driven MI state in the vicinity of $U=\frac{2}{3}\omega$,
as shown in Fig.~\ref{fig:2}(d). The obtained anticrossing is a
result of the second-order process whereby two particles of an MI
state hop to the site of their common neighbor, producing a triply
occupied site and resulting in a state of energy $E=3U$. This process
has been analyzed in Ref.~\citep{Eckardt2008} using the DPT approach.
Plugging the expressions (\ref{eq:Hm}) into Eq.~(\ref{eq:WD2-H-DPT})
one finds that for the resonance condition $3U=q\omega$, the element
$c_{3}\equiv\bra 3\hat{w}_{{\rm D}}^{(2)}\ket{{\rm MI}}$ vanishes
for $q$ odd ($\ket 3$ denotes any one of the states with a triply
occupied site reachable starting from $\ket{\text{MI}}$ in two hops).
Meanwhile, for $q=2$, the strongest coupling is observed for $F/\omega\approx2$.
The quasienergy of the driven MI state is indeed calculated correctly
using DPT, as shown in Fig.~\ref{fig:2}(d). However, as $U$ is
tuned away from resonance, the EDPT yields more accurate results.

\begin{figure*}
\begin{centering}
\includegraphics{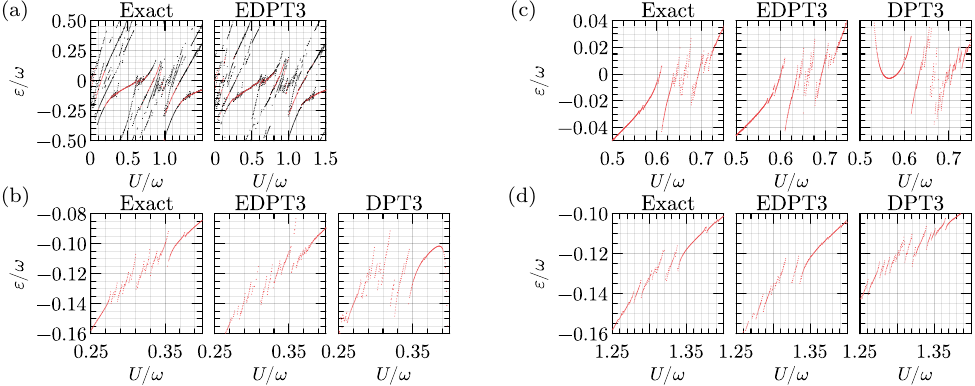}
\par\end{centering}
\caption{\label{fig:3}Assessment of accuracy of DPT and EDPT for an eight-particle
BH system on a $2\times4$ lattice with $F/\omega=2$, $\omega/J=20$.
(a) Quasienergy spectrum for $U\in[0,1.5\omega]$. Quasienergies of
five Floquet modes having the largest overlap with the MI state are
displayed; quasienergies of the mode with the maximum overlap are
highlighted in red. (b) Quasienergies of the driven MI state in the
vicinity of $U=\frac{1}{3}\omega$. (c) Quasienergies of the driven
MI state in the vicinity of $U=\frac{2}{3}\omega$. (d) Quasienergies
of the driven MI state in the vicinity of $U=\frac{4}{3}\omega$.}
\end{figure*}
As noted in Section \ref{subsec:DPT}, the EDPT and DPT approaches
are not equivalent, therefore, the value of $c_{3}$ depends on which
method is used to construct the effective Hamiltonian. These values
are compared in Fig.~\ref{fig:2}(c) for $U=\frac{2}{3}\omega$ and
a range of coupling strengths $F/\omega$. The curves are quite different
in nature: for example, at $F/\omega\approx3.4$ the DPT curve crosses
the zero, while the EDPT curve approaches a local maximum. Vanishing
coupling indicates disappearance of the corresponding anticrossing
in the quasienergy spectrum \citep{Eckardt2008}, which is indeed
the case for $F/\omega\approx3.4$, as confirmed by an exact calculation
(not shown). Thus, even though EDPT yields quasienergies with higher
accuracy than DPT, the matrix elements of $\hat{W}_{{\rm D}}$ constructed
using EDPT do not have such a straightforward interpretation compared
to the case when DPT is used. Another difference between EDPT and
DPT concerns the higher-order terms of the effective Hamiltonian.
In the DPT case, the third-order term $\hat{w}_{{\rm D}}^{(3)}$ gives
an insignificant correction to the second-order theory at $U=\frac{2}{3}\omega$.
The additional processes appearing in the third-order are the creation
of a state featuring three doubly occupied sites (the corresponding
matrix element is $\eta=9.2$ times smaller than $c_{3}$) and the
process of creating a state with a triply occupied site in three hops
($\eta=25$). In the EDPT case, on the other hand, inclusion of the
third-order term $\hat{W}_{{\rm D}}^{(3)}$ gives a substantial improvement
in terms of accuracy.

\begin{figure*}
\begin{centering}
\includegraphics{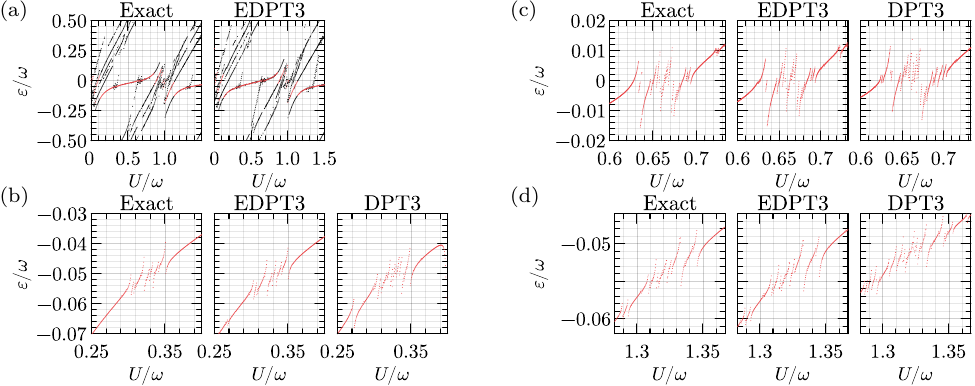}
\par\end{centering}
\caption{\label{fig:4}Same as in Fig.~\ref{fig:3} for $\omega/J=30$.}
\end{figure*}
It is instructive to consider the values of the coupling ratios (\ref{eq:r})
for EDPT. Exactly on resonance $U=\frac{2}{3}\omega$, we find $r_{\max}=0.30$,
which is quite large. However, this value comes from the coupling
of a state corresponding to $\varepsilon_{\alpha m}^{(0)}=-\frac{1}{3}\omega$
with a state corresponding to $\varepsilon_{\alpha'm'}^{(0)}=-\frac{2}{3}\omega$,
therefore, this coupling does not directly influence the driven MI
state, whose reduced energy is $\varepsilon_{*}^{(0)}=0$. The said
coupling is indicated in red in Fig.~\ref{fig:2}(b) depicting the
diagonal elements of $\bar{Q}'$ at $U=\frac{2}{3}\omega$. On the
other hand, all of the degenerate states sharing the value $\varepsilon_{*}^{(0)}=0$
are coupled weakly to states outside the central FZ, as indicated
by the blue numbers in Fig.~\ref{fig:2}(b) (see figure caption for
details). This explains the fact that accurate results have been obtained
using EDPT despite the condition (\ref{eq:r}) not being satisfied.
We remind that the couplings between the states in the central FZ
{[}see green arrows with numbers in Fig.~\ref{fig:2}(b){]} are taken
into account exactly in the EDPT framework. Meanwhile, the couplings
of degenerate states with the nondegenerate ones are treated perturbatively
in the DPT, which explains why the reported DPT results are less accurate
even exactly on resonance.

Let us also discuss the origin of the discontinuity at $U=\frac{5}{8}\omega$
in the EDPT results seen in Fig.~\ref{fig:2}(d). For definiteness,
consider the group of levels characterized by $\varepsilon_{\alpha}^{(0)}=-\frac{1}{3}\omega$
and reduction number $a=3$, appearing in red in Fig.~\ref{fig:2}(b),
where $U=\frac{2}{3}\omega$ is assumed. These are the levels arising
from the unperturbed states of energy $E=4U$. As $U$ decreases,
these levels shift downwards, reaching the lower FZ boundary when
$U$ attains the value of $\frac{5}{8}\omega$. Decreasing $U$ still
further makes this particular group of levels leave the FZ, and another
one (appearing in green in the figure) enters from above. However,
the reduction number for the levels of the latter group is $a=2$.
Since the reduction number directly influences the strength of coupling
with other levels, the coupling with the states under consideration
changes abruptly as $U$ crosses the value of $\frac{5}{8}\omega$.
Although processes such as this one reduce the accuracy of the EDPT,
their impact is not that noticeable if the states of actual interest
are not strongly coupled to the states crossing the boundaries of
FZ. This is certainly the case presently: Our main focus is on the
driven MI state, which is not directly coupled to the $4U$ states.
Consequently, the artifactual discontinuity in Fig.~\ref{fig:2}(d)
is an order of magnitude narrower than the width of the anticrossing
that is of actual interest.

The last case for this parameter set is studied in Fig.~\ref{fig:2}(e)
that displays the quasienergy of the driven MI state in the vicinity
of $U=\frac{4}{3}\omega$. The anticrossing is again a manifestation
of the second-order process producing a triply occupied site. Despite
the width of the anticrossing making up only $\sim2\%$ of the width
of the FZ, it can be calculated accurately using perturbation theory.
Again, EDPT is more accurate than DPT, although the former one yields
a number of false anticrossings.

\subsection{Study case 2: Two-dimensional lattice}

We now turn to a BH system described by the same parameter values
($F/\omega=2$, $\omega/J=20$), but this time on a $2\times4$ lattice
periodic in the $x$-direction (we direct the $x$-axis along the
longer dimension of the lattice). The quasienergy spectrum calculated
for $U\in[0,1.5\omega]$ is shown in Fig.~\ref{fig:3}(a). We notice
that the curve of the driven MI state undergoes many more anticrossings
compared to the above case of a one-dimensional lattice, indicating
richer system dynamics. The EDPT results are less accurate here, featuring
erroneous and noticeable discontinuities. Nevertheless, the quasienergies
can be considered calculated qualitatively correctly in the vicinity
of the main anticrossing at $U=\omega$.

According to Eq.~(\ref{eq:Ht}), driving renormalizes the hopping
strength only for hopping along the $x$-axis. As a result, the DPT
theory no longer predicts that $c_{3}$ vanishes for all $F$ when
$U=\frac{m}{3}\omega$ with $m$ odd. Indeed, Fig.~\ref{fig:3}(b)
confirms that numerous anticrossings appear in the vicinity of $U=\frac{1}{3}\omega$.
Their presence is predicted by both EDPT and DPT, albeit the exact
results are matched only qualitatively. Meanwhile, in the cases $U=\frac{2}{3}\omega$
and $U=\frac{4}{3}\omega$ the EDPT ensures higher accuracy, providing
a considerable improvement over DPT {[}see Figs.~\ref{fig:3}(c)
and (d){]}.

Finally, we consider the results obtained for the case of a higher
driving frequency: $\omega/J=30$. As expected, the accuracy of the
perturbation theory is higher in this case, which is confirmed by
the plots in Fig.~\ref{fig:4}. Figure \ref{fig:4}(a) shows that
EDPT yields accurate results on a large scale, capturing all the essential
features of the exact quasienergy spectrum. The close-up views of
the spectrum in the vicinity of $U=\frac{1}{3}\omega$, $U=\frac{2}{3}\omega$
and $U=\frac{4}{3}\omega$, shown in Figs.~\ref{fig:4}(b), (c),
and (d), respectively, display that EDPT is capable of providing quantitatively
correct results in this regime. The DPT is certainly applicable as
well, although the accuracy is lower.

\subsection{Computational cost of the methods}

Concluding the discussion of accuracy of the considered methods, let
us compare the associated computational costs. Calculation of quasienergies
using both DPT and EDPT comes down to a Hermitian matrix diagonalization.
In the EDPT case, one is required to diagonalize the matrix whose
size is given by the total number of states of the unperturbed systems.
In the studied eight-particle systems, there are $N=6435$ such states.
In the DPT case, the size is given by the number of states sharing
the same value of reduced energy $\varepsilon_{*}^{(0)}$. For example,
on a $2\times4$ lattice and at $U=\frac{2}{3}\omega$, there are
2017 states sharing the value of $\varepsilon_{*}^{(0)}=0$ {[}cf.
Fig.~\ref{fig:2}(b){]}. Both methods thus suffer from exponential
growth of the required computation resources, which can only be alleviated
by limiting the number of states taken into consideration.

The numerically exact calculation of the quasienergies via the single-period
evolution operator $\hat{P}$ requires performing a (unitary) $N\times N$
matrix diagonalization, similarly to the the EDPT case. However, the
construction of $\hat{P}$ additionally requires propagating the Schr\"{o}dinger
equation for one period of the drive $N$ times (for each basis state
as the initial condition). Although the specific time required to
solve the differential equations depends on multiple factors (such
as the solver, required tolerance, and hardware), in our experience
\citep{FloquetSystems.jl} the exact calculation of the quasienergies
for a single value of $U$ took 2.5--6 times longer than the EDPT
calculation and, importantly, required $\sim15$ times more memory
(see Appendix B for details).

\section{Conclusion}

Let us now summarize the results. We have provided an extension of
the conventional degenerate perturbation theory that improves the
accuracy of the calculation of quasienergy spectra of periodically
driven systems. Application of the theory to the driven Bose--Hubbard
system has shown that third-order EDPT yields results matching the
exact ones on a quantitative level. While the exact applicability
criterion is difficult to formulate, the simple condition (\ref{eq:r})
together with the provided example calculations may serve as a basis
for predicting the expected accuracy. Generally, the accuracy is expected
to improve with increasing driving frequency.

Along the way, we have also studied the application of the conventional
DPT. It has an advantage that the matrix elements of the resulting
effective Hamiltonian $\hat{w}_{{\rm D}}^{[n]}$ admit a straightforward
interpretation. Specifically, it enables one to make qualitative predictions
about the possible appearance of the anticrossings in the quasienergy
spectrum, and the matrix elements of $\hat{w}_{{\rm D}}^{[n]}$ may
be used to estimate the relative probabilities of various resonant
excitation processes \citep{Eckardt2008,Weinberg2015,Eckardt2015}.
On the other hand, since $\hat{w}_{{\rm D}}^{[n]}$ consists of two
decoupled blocks (describing Hilbert spaces $\mathbb{D}^{0}$ and
$\mathbb{D}^{1}$, cf. Section \ref{subsec:DPT}), which is clearly
a bold approximation, it might not capture important properties of
the system, such as the existence of a Floquet dynamical symmetry
\citep{Koki2020,Sarkar2022a,Sarkar2022b}. In this respect, the effective
Hamiltonian $\hat{W}_{{\rm D}}^{[n]}$ provided by EDPT is expected
to be more useful: existence of an operator $\hat{A}$ such that $[\hat{W}_{{\rm D}}^{[n]},\hat{A}]=\lambda\hat{A}$
with $\lambda$ a real number would imply the existence of a dynamical
symmetry in the effective system \citep{Koki2020,Buca2019}, to the
$n$th order of the perturbation theory. This might help in exploring
the platforms for constructing discrete time crystals.

The performed analysis has shown that the accuracy of the quasienergies
obtained using DPT is considerably lower than that of EDPT. Moreover,
while DPT is applicable only in the vicinity of resonances, EDPT remains
equally useful if the driving is not resonant. The advantages of EDPT,
however, come at the expense of increased numerical effort since the
resulting effective Hamiltonian is of the same size as the unperturbed
one. For large systems, its diagonalization becomes prohibitively
costly, and one might need to reduce the number of considered states
by excluding highly-excited ones, for example. Another option is to
adopt a scheme intermediate between DPT and EDPT so that only some
of the couplings between the states in the FZ are treated exactly,
and others are taken into account perturbatively. The required formalism
has been presently provided.
\begin{acknowledgments}
Authors would like to thank Andr\'{e} Eckardt for useful discussions.
This work was supported by the Research Council of Lithuania, Grant
no. S-LJB-24-2.

Calculations have been performed using a number of software packages
\citep{JuliaDiffEq,JuliaDiffEq2,JuliaDiffEqTsit} written in Julia
\citep{Julia}.

\newpage{}
\end{acknowledgments}

\onecolumngrid

\section*{Appendix A: Third-order expressions for the effective Hamiltonian}

\setcounter{equation}{0}
\renewcommand{\theequation}{A\arabic{equation}}Here, we provide the third-order expressions for the effective Hamiltonians.
In the DPT framework, one obtains

\begin{equation}
\begin{split}\bra{\alpha'}\hat{w}_{{\rm D}}^{(3)}\ket{\alpha} & =\frac{1}{2}\delta_{A'A}\sum_{\beta,\gamma}\sum_{p\neq0\text{ if }C=A}\left[\delta_{AB}\frac{1}{\varepsilon_{\beta}-\varepsilon_{\gamma}-p\omega}\right.\\
 & \times\left(\frac{\bra{\alpha'}\hat{H}_{b-a'}\ket{\beta}\bra{\beta}\hat{H}_{c-b-p}\ket{\gamma}\bra{\gamma}\hat{H}_{a-c+p}\ket{\alpha}}{\varepsilon_{\gamma}-\varepsilon_{\alpha'}+p\omega}+\frac{\bra{\alpha'}\hat{H}_{c-a'-p}\ket{\gamma}\bra{\gamma}\hat{H}_{b-c+p}\ket{\beta}\bra{\beta}\hat{H}_{a-b}\ket{\alpha}}{\varepsilon_{\gamma}-\varepsilon_{\alpha}+p\omega}\right)\\
 & +\delta_{BC}\bra{\alpha'}\hat{H}_{b-a'-p}\ket{\beta}\bra{\beta}\hat{H}_{c-b}\ket{\gamma}\bra{\gamma}\hat{H}_{a-c+p}\ket{\alpha}\\
 & \left.\times\left(\frac{1}{(\varepsilon_{\alpha'}-\varepsilon_{\beta}-p\omega)(\varepsilon_{\alpha'}-\varepsilon_{\gamma}-p\omega)}+\frac{1}{(\varepsilon_{\alpha}-\varepsilon_{\beta}-p\omega)(\varepsilon_{\alpha}-\varepsilon_{\gamma}-p\omega)}\right)\right]\\
 & -\frac{1}{12}\delta_{A'A}\sum_{\beta\gamma}\sum_{p\neq0\text{ if }A=B}\sum_{\begin{subarray}{c}
q\neq p\text{ if }B=C\\
q\neq0\text{ if }C=A
\end{subarray}}\bra{\alpha'}\hat{H}_{b-a'-p}\ket{\beta}\bra{\beta}\hat{H}_{c-b+p-q}\ket{\gamma}\bra{\gamma}\hat{H}_{a-c+q}\ket{\alpha}\\
 & \times\left[\frac{3}{\varepsilon_{\beta}-\varepsilon_{\alpha}+p\omega}\left(\frac{1}{\varepsilon_{\alpha}-\varepsilon_{\gamma}-q\omega}+\frac{1}{\varepsilon_{\beta}-\varepsilon_{\gamma}+(p-q)\omega}\right)\right.\\
 & \qquad+\frac{3}{\varepsilon_{\alpha'}-\varepsilon_{\gamma}-q\omega}\left(\frac{1}{\varepsilon_{\beta}-\varepsilon_{\gamma}+(p-q)\omega}+\frac{1}{\varepsilon_{\beta}-\varepsilon_{\alpha'}+p\omega}\right)\\
 & \qquad+\frac{1}{\varepsilon_{\beta}-\varepsilon_{\gamma}+(p-q)\omega}\left(\frac{1}{\varepsilon_{\beta}-\varepsilon_{\alpha'}+p\omega}+\frac{1}{\varepsilon_{\alpha}-\varepsilon_{\gamma}-q\omega}\right)\\
 & \qquad\left.+\frac{2}{(\varepsilon_{\alpha}-\varepsilon_{\gamma}-q\omega)(\varepsilon_{\beta}-\varepsilon_{\alpha'}+p\omega)}\right]
\end{split}
\label{eq:WD3}
\end{equation}
Here, integers $A$, $A'$, $B$, and $C$ indicate which subspaces
the states belong to: $\ket{\alpha}\in\mathbb{D}^{A}$, $\ket{\alpha'}\in\mathbb{D}^{A'}$,
$\ket{\beta}\in\mathbb{D}^{B}$, $\ket{\gamma}\in\mathbb{D}^{C}$.
In the EDPT framework, all states are assigned to the same subspace,
therefore, in that case one should put $A=A'=B=C$ in the above equation.

\bigskip{}

\twocolumngrid

\section*{Appendix B: Benchmarks of the methods}

In this Appendix we provide benchmarks of the methods and additional
technical details.

\begin{figure}[H]
\centering{}\includegraphics{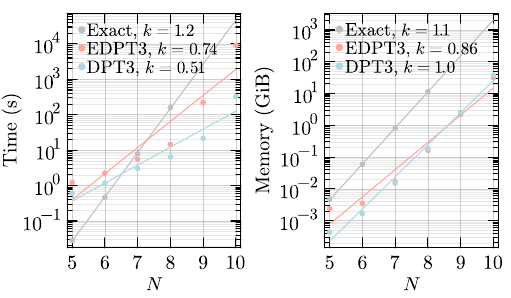}\caption{\label{fig:5}Scaling of computation time and required memory with
the system size. The code \citep{FloquetSystems.jl} was benchmarked
on a Mac mini computer with Apple M2 Pro chip (8 performance cores
variant) and 32\,GB RAM. The exact calculations for systems with
$N>8$ were not performed as the memory requirements exceeded resources
available on the test machine. The reported memory utilization was
measured as the total amount of memory allocated during the calculation.
This was chosen as a reproducible criterion that reflects the scaling
behavior. The actual amount of memory required is highly dependent
on the implementation details.}
\end{figure}

The scaling of the computational time and required memory with the
system size is shown in Fig.~\ref{fig:5}. The performed calculations
correspond to those presented in Fig.~\ref{fig:2} at a single point
of $U=\frac{2}{3}\omega$, for lattice sizes $N=5$ through 10 (with
unit filling). The data is depicted on a semi-logarithmic scale, together
with an exponential fit $y=10^{kN}+y_{0}$; the coefficients $k$
are provided in the legends. It is apparent that the EDPT provides
a noticeable advantage in terms of calculation times compared to the
exact approach and requires more than an order of magnitude less memory.
The memory requirements for DPT are similar to those of EDPT because
we were constructing the effective Hamiltonian $\hat{w}_{{\rm D}}^{[3]}$
for both the degenerate and nondegenerate subspaces, although we were
diagonalizing only the block corresponding to the degenerate space.

In practice, one often needs to repeat the calculation of quasienergies
for a range of one or more system parameters, as was done in our analysis
in the main text.
 The loop scanning over the parameters can be easily parallelized
on a multicore machine or a cluster, but the memory requirements might
put a limit on the number of processes that can be run in parallel.
For example, the exact calculation shown in Fig.~\ref{fig:3}(a),
where $\sim300$ values of $U$ were scanned, was performed using
only 21 of 64 cores of an AMD 2990WX CPU because that already used
almost all available RAM space (121\,GB out of 128\,GB). The execution
time was 6 hours. Meanwhile, EDPT3 allowed us to utilize all 64 cores
while requiring 22\,GB RAM in total; the calculation finished in
50 minutes.

\end{document}